\documentclass{Interspeech2024}
\usepackage{multirow}
\usepackage{bbding}
\usepackage{makecell}
\usepackage{graphicx} 
\usepackage[table]{xcolor}
\interspeechcameraready
\title{ED-sKWS: Early-Decision Spiking Neural Networks for Rapid, and Energy-Efficient Keyword Spotting
}

\name[affiliation={1}]{Zeyang}{Song}
\name[affiliation={1}]{Qianhui}{Liu$^*$}
\name[affiliation={1}]{Qu}{Yang}
\name[affiliation={1}]{Yizhou}{Peng}
\name[affiliation={1,2}]{Haizhou}{Li}

\address{$^1$National University of Singapore, Singapore\\
$^2$School of Data Science, Shenzhen Research Institute of Big Data,\\ The Chinese University of Hong Kong, Shenzhen, Guangdong, 518172, P.R. China}
\email{zeyang\_song@u.nus.edu, qhliu@nus.edu.sg}


\begin{document}

\maketitle

\begin{abstract}
Keyword Spotting (KWS) is essential in edge computing requiring rapid and energy-efficient responses. Spiking Neural Networks (SNNs) are well-suited for KWS for their efficiency and temporal capacity for speech. To further reduce the latency and energy consumption, this study introduces ED-sKWS, an SNN-based KWS model with an early-decision mechanism that can stop speech processing and output the result before the end of speech utterance. Furthermore, we introduce a Cumulative Temporal (CT) loss that can enhance prediction accuracy at both the intermediate and final timesteps. To evaluate early-decision performance, we present the SC-100 dataset including 100 speech commands with beginning and end timestamp annotation. Experiments on the Google Speech Commands v2 and our SC-100 datasets show that ED-sKWS maintains competitive accuracy with 61\% timesteps and 52\% energy consumption compared to SNN models without early-decision mechanism, ensuring rapid response and energy efficiency.
\end{abstract}

\renewcommand{\thefootnote}{}
\footnotetext{This work was supported in part by Shenzhen Science and Technology Research Fund (Fundamental Research Key Project Grant No. JCYJ20220818103001002), Shenzhen Science and Technology Program ZDSYS20230626091302006 and IAF, A*STAR, SOITEC, NXP and National University of Singapore under FD-fAbrICS: Joint Lab for FD-SOI Always-on Intelligent \& Connected Systems (Award I2001E0053). }
\footnotetext{$^*$ Corresponding author. }

\section{Introduction}
Keyword Spotting (KWS) refers to the task of identifying specific words or phrases within audio streams, necessitating both rapid and accurate recognition. This capability is foundational to numerous applications in edge computing, such as smartphones, smart speakers, and in-vehicle audio systems, where priorities include rapid real-time responses, energy efficiency, and high accuracy \cite{lopez2021deep,song2022knowledge}.

Spiking Neural Networks (SNNs) are the third generation of neural networks that more closely mimic the behavior of biological neural networks in the brain \cite{wu2018spiking,yin2021accurate,liu2020unsupervised, Liu2024120660}. Unlike traditional neural networks that use continuous values for activation, SNNs use discrete events (spikes) for information communication. Within an SNN, neurons remain inactive until their membrane potential surpasses a specific threshold. This event-driven mechanism endows SNNs with superior energy efficiency, making them especially suitable for KWS tasks \cite{wu2018biologically,yilmaz2020deep,yang2022deep,sun2023adaptive, song2024spiking}. Furthermore, the inherent temporal dynamics of spiking neurons align closely with the temporal nature of speech, rendering them inherently well-suited for speech processing \cite{bellec2018long,yin2021accurate, yang2024svad}. Considering the timestep-by-timestep property of feed-forward SNN, similar to vanilla RNNs, SNNs are suitable for real-time KWS \cite{choi2019temporal}, which needs to process the input speech frame by frame.

Several current SNN methods, like \cite{wu2018biologically, yilmaz2020deep, yang2022deep}, generally require the complete input samples before making a final decision (called late-decision in this paper), thereby not leveraging SNNs' inherent ability for real-time processing to its fullest. To harness SNNs' real-time capabilities for rapid response and energy efficiency to their maximum, an early-decision mechanism has been incorporated into SNN models. This mechanism enables models to terminate processing and deliver outcomes at intermediate frames, facilitating significant energy conservation and improved response speed. However, this approach may lead to potential information loss due to premature data at the point of early decision. This trade-off highlights the need for training techniques that maintain high accuracy for early decision.

\begin{figure}[t]
  \centering
  \includegraphics[width=\linewidth]{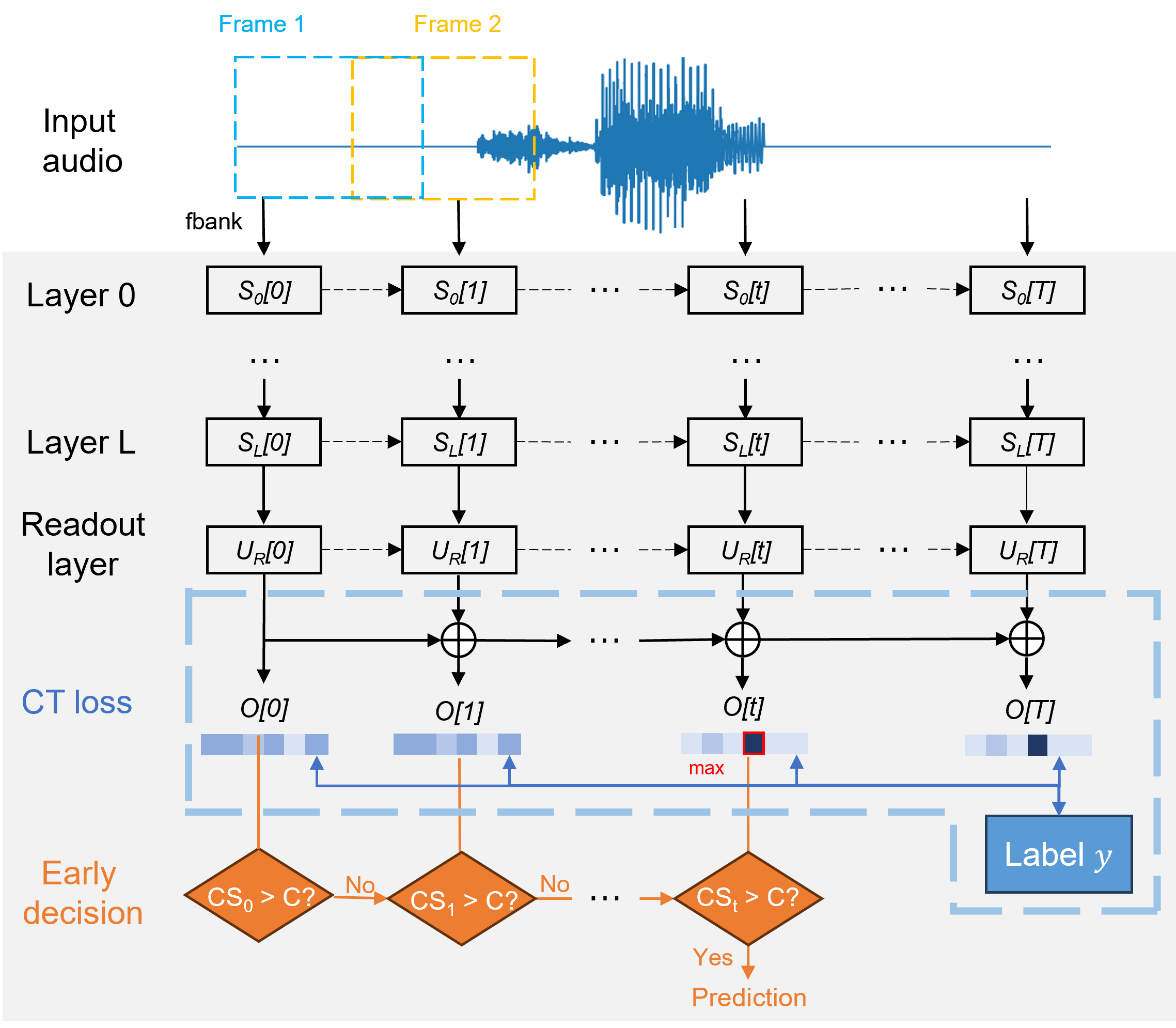}
  \caption{The overall diagram of ED-sKWS with CT loss and early decision mechanism. The ED-sKWS system processes the fbank features of the input audio using a feed-forward SNN, which is indicated by the grey box. In the feedforward SNN, we feed frame $t$ to timestep $t$ of SNN, processing the data frame-by-frame. With an early-decision mechanism, the ED-sKWS ends processing and delivers the output once the confidence score ($CS$) exceeds a predefined threshold ($C$). 
  }
  \label{fig:diagram}
\end{figure}

This paper proposes an early-decision SNN-based model for KWS, referred to as ED-sKWS (as shown in Fig. \ref{fig:diagram}). This model is grounded in a fully-connected SNN architecture for real-time processing, coupled with a proposed Cumulative Temporal (CT) loss tailored for early-decision training. 
By feeding the feature of each speech frame into the corresponding timestep of SNN, the model can process speech information in a frame-synchronized manner, thus facilitating real-time processing and immediate response. The CT loss is proposed to enhance the prediction performance at early-decision timesteps, considering not only the current frame information but also the comprehensive contribution of historical information. This dual consideration allows for a more complete understanding and use of temporal information.

Existing KWS datasets, like Google Speech Commands \cite{speechcommandsv2}, adopt a standardized format by padding extracted utterances with zeros to ensure a consistent sample length. However, this poses a challenge in our early decision study as we cannot know the specific beginning and end time of the keyword, which complicates the accurate assessment of response speed. Even though we can consider the beginning and end of the sample as the beginning and end time of the keyword, it will lead to considerable inaccuracies. To overcome this limitation and offer an extensive KWS dataset for model evaluation, we present SC-100 including 100 speech commands with detailed timestamp annotations for the beginning and end timestamps of the commands. These annotations enable precise tracking of the command's actual occurrence, serving not only to quantify the benefits of early decision but also to be applied to other related speech processing applications, such as Voice Activity Detection (VAD).

The main contributions can be summarized as follows:
\begin{itemize}
    \item We introduce ED-sKWS, an early-decision SNN-based framework designed for fast, energy-efficient, and accurate keyword spotting in real time.
    \item We propose a novel Cumulative Temporal (CT) loss for enhancing the early decision capabilities of SNNs. 
    \item We develop a large-scale keyword spotting dataset SC-100 for early-decision evaluation, including 100 commonly used speech commands relevant to daily life. This dataset is characterized by the inclusion of precise beginning and end timestamps for each command, facilitating a more accurate and detailed analysis of keyword detection and timing.
\end{itemize}

\section{Methods}

\subsection{SNN for real-time keyword spotting}

To fully harness the capabilities of SNNs for real-time KWS, our ED-sKWS model, designed for real-time keyword spotting, utilizes a feed-forward fully connected SNN. Unlike the Spiking Convolutional Neural Network (SCNN) approaches \cite{yilmaz2020deep, yang2022deep} that treat the spectrogram of input audio as an image, requiring additional timesteps for processing, our fully connected SNNs process a single frame of the spectrogram at one timestep. This process eliminates the need for extra timesteps and synchronizes the SNN model's timesteps with the audio utterances' time dimension, akin to Recurrent Neural Networks (RNNs). This frame synchronization facilitates predictions at every timestep, enabling real-time prediction and the potential for early decision.

To further enhance the temporal capacity of our model, we use adaptive Leaky Integrate-and-Fire (LIF) neurons similar to \cite{bittar2022surrogate} as the spiking neuron. Our choice of the adaptive LIF neuron model, a variant of the standard LIF model \cite{gerstner2002spiking}, is motivated by its adaptive mechanisms that enhance control over information flow within the network. The adaptive LIF model's dynamics are described by the following equations:

\begin{align}
    &I_l[t] = \beta * \Sigma_i w_iS_{l-1}[t-1] + a U_l[t-1] + b S_l[t-1]\\
    &U_l[t] = \alpha * (U_l[t-1] - V_{th}S_l[t-1]) + I_l[t] \\
    \label{LIF_thres} &S_l[t] = \mathbb{H}(U_l[t] - V_{th})
\end{align}
where $S_{l-1}[t-1]$ represents the input spike from layer $l-1$ at timestep $t-1$. The terms $I_l[t]$ and $U_l[t]$ represent the adaptive synaptic current and the membrane potential in layer $l$ at timestep $t$, respectively. The constants $\alpha$ and $\beta$ determine the information decays within the neuron. The adaptation mechanism is further controlled by parameters $a$ and $b$, which are associated with subthreshold dynamics and spike-triggered responses. The firing of the neuron is governed by a Heaviside step function, as denoted in Eq. \ref{LIF_thres}. When the membrane potential $U[t]$ surpasses the firing threshold $V_{th}$, an output spike is generated.

\subsection{Cumulative Temporal (CT) loss}

In this section, we describe the proposed training loss for our ED-sKWS model. In previous studies, several loss functions like the average spike-rate based loss \cite{wu2018spatio} and cumulative loss \cite{bittar2022surrogate} are used in various late-decision SNNs. They only compute the loss given the output in the last timestep, failing to optimize the output in the intermediate timesteps. TET loss \cite{deng2022temporal} solved this problem by optimizing outputs in each isolated timestep, and proved to be effective in the static dataset where inputs in all timestep are identical. However, it does not perform well in continuously varying data like speech commands, as the TET loss fails to model the historical information.  Therefore, we propose the Cumulative Temporal (CT) loss, which considers not only the current frame information but also the comprehensive contribution of historical information, allowing for a more complete understanding and use of temporal information The CT loss is described as:

\begin{align}
    &O[t] = \sum^{t}_{i=0} softmax(U_R [i])\\
    &L_{CT} = \frac{1}{T} \sum^{T}_{t=0} L_{CE}[O[t], y]
\end{align}

where $O[t]$ denotes the cumulative historical information up to the current timestep, expressed as a softmax cumulative sum of potential over time. By optimizing $O[t]$ at each timestep, the cumulative output aligns more closely with the target distribution, facilitating accurate predictions in the intermediate timestep. 

\subsection{Early decision method}

In addition to the CT loss, an early-decision mechanism is essential for the ED-sKWS model to determine the optimal stopping point. We utilize the confidence score, denoted as $CS_t = \max(\text{softmax}(O[t]))$, as a proxy for this purpose. This score quantifies the model's confidence in classifying an input pattern into one of the object classes—the higher the $CS_t$, the greater the model's confidence in its prediction. An early-decision threshold, $C$, is established: when $CS_t$ surpasses $C$, the model deems its prediction sufficiently reliable to halt further processing. Conversely, if $CS_t$ is below this threshold, the model persists in integrating information from the following frames until it reaches a level of confidence to make a decision.

To effectively assess the performance of early-decision SNN models in KWS, we utilize two key metrics: early-decision accuracy ($Acc^t$) and late-decision accuracy ($Acc^T$). $Acc^t$ evaluates predictions made during the early-decision phase, whereas $Acc^T$ measures the accuracy based on the output at the final timestep.

\section{SC-100 Dataset}


In this study, we present the SC-100 dataset, specifically designed for the evaluation of early-decision performance in KWS systems. For an accurate assessment of response speed, we need to know when the end time of keywords in each sample. However, the widely used Google Speech Command dataset \cite{speechcommandsv2} offers keyword samples without temporal annotations indicating the end of keyword utterances. To address this gap, we develop a large-scale SC-100 dataset including 100 speech commands in our daily life, along with detailed timestamps annotations for the beginning and end time of each command within the audio samples, aiming to enhance the precision of early-decision assessments in KWS applications.

To develop SC-100 dataset, we first employ the KeywordMiner tool \cite{meneses2022sidi} to extract the words from the LibriSpeech dataset \cite{panayotov2015librispeech}. The KeywordMiner consists of an aligner to obtain the word and the corresponding timestamp in the sentence samples and a segmenter to export the words from sentences in LibriSpeech. However, the raw output generated by KeywordMiner is not immediately suitable for direct use in KWS systems. First, not all segmented words can be considered keywords because their meanings may be inconsistent with the functional requirements of the KWS system, such as "a" or "to"; second, the duration of some word utterances is too short to be used as training or testing samples. To improve the quality of the dataset, we select 100 keywords that are representative and have a duration of more than 0.4 seconds. 

The selected 100 keywords are partitioned to reflect the multifaceted scenarios of everyday life:
\begin{itemize}
    \item \textbf{Smart home}: change, turn, set, on, off, light, white, dark, up, down, slowly, open, close, door, window.
    \item \textbf{Entertainment}: show, point, end, next, first, second, last, back, again, second, voice, quiet, black, blue, white, red, green, play, ready, rest. 
    \item \textbf{Alarm}: set, stop, change.
    \item \textbf{Robot Helper}: go, return, come, stay, left, right, there, here, forward, in, out, straight, wait, hold, put, above, below, before, after, north, south, outside, top, find, give, make, get, leave.
    \item \textbf{Office}: book, use, table, room, time, hour, minute, work, zero, one, two, three, four, five, six, seven, eight, nine, office.
    \item \textbf{Self assistant}: call, send, speak, talk, read, take, bring, box, later, remember, tell, water, morning, afternoon, evening.
\end{itemize}

The SC-100 dataset contains 313,951 keyword utterances, with each class containing between 1,000 and 4,000 utterances. The potential class imbalances can be addressed by applying uniform sampling weights across all classes to facilitate fair training conditions. The extracted utterances vary in duration from 0.4 to 1 second. To offer a standardized format, we created 1-second audio samples by padding the extracted utterances with a random number of zeros at the beginning and end, similar to the Google Speech Commands dataset. However, it introduces a challenge in our early decision study or other related tasks: we cannot know the specific beginning and end time of the keyword, which complicates the accurate assessment of response speed. To address this issue, we meticulously annotate the beginning and end times of each keyword utterance within the samples. These timestamp annotations are crucial for precisely validating the early-decision mechanism, as detailed in Section 4.4. Additionally, these annotations are expected to facilitate the advancement of related technologies, such as VAD models.

\section{Experiment results}

\subsection{Experiment setup}
In this section, we evaluate the KWS performance on Google Speech Command Dataset V2 \cite{speechcommandsv2}, which contains 105,829 one-second utterances of 35 commands, and our proposed SC-100 dataset. The sampling rate of these two dataset is 16kHz. 


For preprocessing, we extract fbank features on the raw audio, using 40 filters with a window length of 25ms. Given that the sample length across both datasets is fixed at 1 second, this results in a total of 98 frames, which is also the number of timesteps in SNN.

In our proposed ED-sKWS model, we use feed-forward SNN with two hidden layers and a readout layer \cite{bittar2022surrogate}. For fair comparison with previous works, our evaluation includes two model configurations: one with 128 hidden neurons (27.63K parameters) and another with 512 hidden neurons (306.80K parameters).

\subsection{Earlier decision time with comparable accuracy}
\label{sec:comparison}
We compared our proposed ED-sKWS with various existing SNN-based KWS models. Table \ref{tab:comparison} demonstrates superior performance with shorter decision time and lower energy consumption. This advantage stems from our early decision mechanism. Taking the example of model with 512 hidden neurons, our ED-sKWS can make predictions approximately 36 timesteps (also 36 frames) before the sample ends, reducing about 38\% inference latency in the KWS system. 

Following \cite{li2021differentiable,rathi2021diet,che2022differentiable}, we estimate our computational energy consumption based on 45nm CMOS technology \cite{horowitz20141}, with MAC operations consuming 4.6pJ of energy and accumulations consuming 0.9pJ. Compared to the adLIF baseline method \cite{bittar2022surrogate}, our ED-sKWS significantly reduces energy consumption by about 48\% while only experiencing a performance decrease of 0.08\%. This capability highlights the potential of our proposed ED-sKWS in KWS applications that demand high performance, rapid response, and energy efficiency.

\begin{table}[]
\centering
\caption{Comparison of existing SNN-based KWS models with the number of parameters, accuracy, decision time $t_d$, and estimated energy consumption.}
\label{tab:comparison}
\resizebox{\linewidth}{!}{\begin{tabular}{lcccc}
\toprule
\textbf{Methods}& \textbf{Params.}& \textbf{Acc. (\%)} & \textbf{$t_d$}& \textbf{Energy($\mu J$)}\\
\midrule

Yilmaz et al.\cite{yilmaz2020deep} &  117K & 75.20 & 98 & -                     \\
MSAT \cite{he2024msat} &  500K  & 87.33 & 98  & -                  \\
sKWS \cite{yang2022deep}&  86.5K  & 91.7 & 98  & 247.52                   \\
Spiking-LEAF \cite{song2024spiking}&  306.80K  & 93.02 & 98   &  29.20   \\
\multirow{2}{*}{adLIF \cite{bittar2022surrogate}}& 27.63K & 90.46& 98  &5.36                    \\
                      & 306.80K   & 93.12    & 98     & 45.41               \\
\midrule
\multirow{4}{*}{ED-sKWS} & \multirow{2}{*}{27.63K}  &90.14     & 66.07 & 2.85                    \\
                      &      &90.52     & 98    &4.74                  \\
                      \cline{2-5} 
                      & \multirow{2}{*}{306.80K}& 93.04    &60.46        &23.68           \\
                      &     & 93.15    & 98    &34.62             \\
\bottomrule
\end{tabular}}
\end{table}

\subsection{Ablation study on different loss functions}
\label{loss ablation}
In this section, we present an ablation study for the effectiveness of our proposed CT loss.  We focus on comparing the average decision time among all samples $\tilde{t_d}$, early-decision accuracy $Acc^t$, late-decision accuracy at the last timestep $Acc^T$ and average spike rate over the whole sample $R^T$ across different types of loss functions. The loss functions for comparison are listed below:
\begin{itemize}
    \item Spike-rate loss: $L_{CE} = L_{CE}(\frac{1}{T}\sum^T_{t=0} U_R [t], y)$
    \item TET loss: $L_{TET} = \frac{1}{T} \sum^{T}_{t=0} L_{CE}(U_R [t], y)$
    \item Cumulative loss: $L_{CE} = L_{CE}(\sum^{t}_{i=0} softmax(U_R [i]), y)$
\end{itemize}
The comparative results are presented in Table \ref{tab:loss_ablation}. This results reveals that the CT loss not only demonstrates superior performance in the final timestep of the sample, where all temporal information is available, but also achieves comparable early-decision accuracy with fewer timesteps. This finding highlights the efficacy of CT loss in optimizing decision-making accuracy both in the early stages and after the input sequence. 



\begin{table}
\centering
\caption{ED-sKWS with different loss functions}
\label{tab:loss_ablation}
\begin{tabular}{ccccc}
\toprule
  \textbf{Loss type} &  \textbf{$Acc^t(\%)$} &  \textbf{$\tilde{t_d}$} & \textbf{$Acc^T(\%)$}\\
\midrule
Spike-rate loss \cite{wu2018spatio} & 88.18 &65.52 & 92.73 \\
TET \cite{deng2022temporal} &     91.24 &      61.24&   91.37\\
Cumulative \cite{bittar2022surrogate} &     93.02 & 66.23 &    93.12\\
CT loss&     93.04 &      60.46 &    93.15\\
\bottomrule
\end{tabular}
\end{table}

\subsection{Experiments on SC-100 dataset}
We also test our ED-sKWS model on our proposed SC-100 dataset with beginning and end annotation for each keyword. A crucial part of our experiments involved comparing the model's early-decision time with the end time of the keyword ($t_{end}$). We use the metric $\Delta t_d = \frac{1}{N} \sum^N (t_d - t_{end})$ to measure the average discrepancy between the model's decision time ($t_d$) and $t_{end}$. This metric evaluates the precision of our model's early stopping ability. As shown in Table \ref{tab:100words}, our model, powered by the proposed CT loss, not only predicts keywords ahead of the end time of the keywords but also outperforms other loss functions in terms of early-decision accuracy $Acc^t$, while maintaining the late-decision performance.

Furthermore, an interesting observation from the SC-100 dataset is the higher early-decision accuracy $Acc^t$ compared to late-decision accuracy $Acc^T$. Unlike the Google Speech Commands dataset, the average end time of keyword utterances in SC-100 dataset is around 62.96 timesteps, which means there will be a long silence after the keyword utterances. This scenario poses a memory challenge for spiking neural networks, as they need to preserve learned features until the final timestep. Here, our early-decision mechanism proves to be beneficial, enabling the model to predict flexibly around the estimated keyword endpoint, thereby mitigating the risk of information loss due to extended silences.

\vspace{2pt}
\begin{table}
\centering
\vspace{2pt}
\caption{Results on SC-100 dataset}
\label{tab:100words}
\begin{tabular}{lcccc}
\toprule
 \textbf{Loss type} &  $Acc^t(\%)$ &  $\tilde{t_d}$ &  $\Delta t_d$ &  $Acc^T(\%)$ \\
\midrule
   Spike-rate loss \cite{wu2018spatio}&90.72 &70.32 &+7.33 &91.76 \\
   TET \cite{deng2022temporal}&91.17 &64.21 &+1.22 &91.20 \\
Cumulative \cite{bittar2022surrogate} & 93.20 &68.28 &+6.32 & 93.07\\
   ED-sKWS &  93.21 &59.11 &  -3.85 & 93.16\\
\bottomrule
\end{tabular}
\end{table}

\subsection{How the early decision reduce energy consumption?}
\begin{figure}[t]
  \centering
  \includegraphics[width=\linewidth]{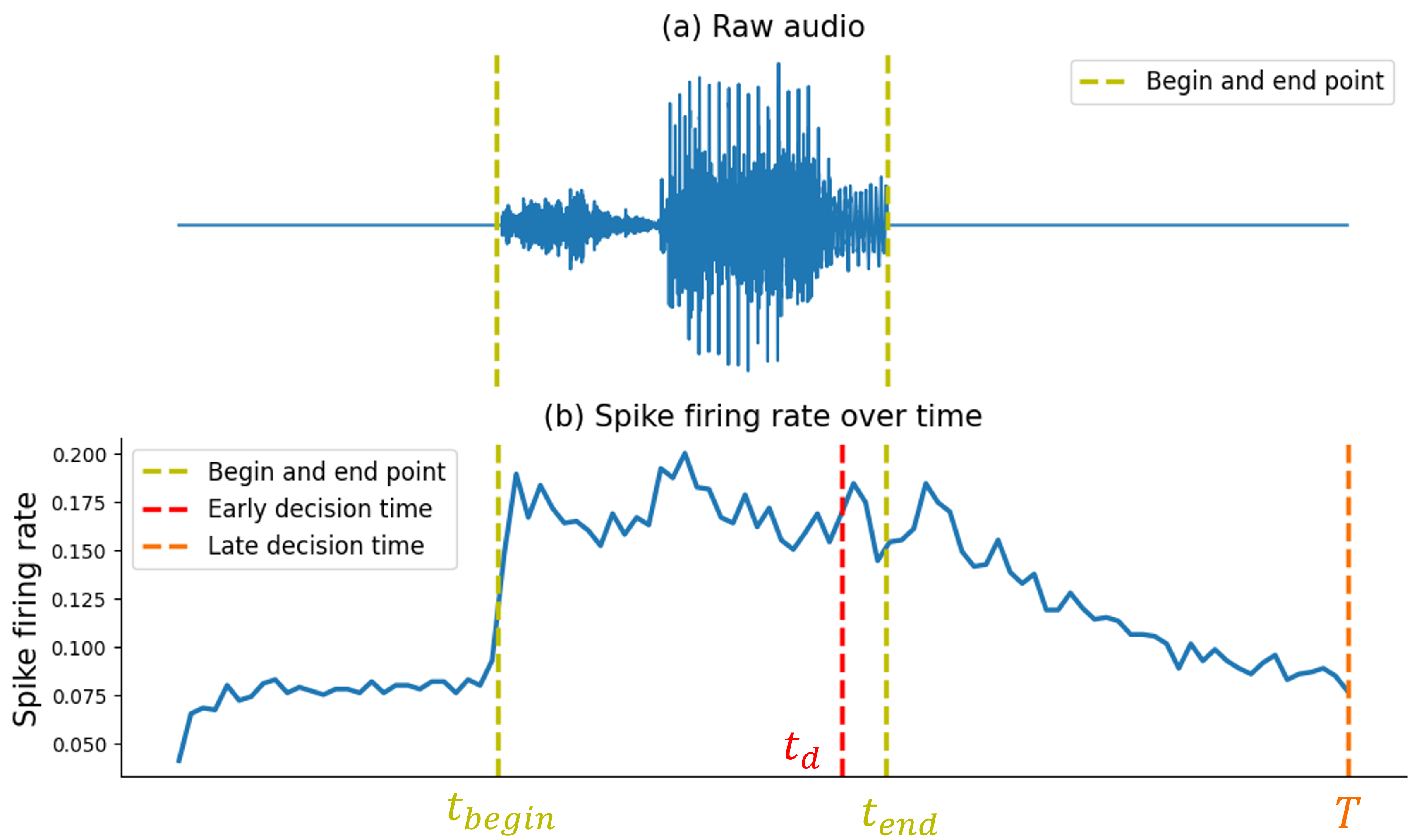}
  \caption{ Visualization of the raw audio signal and the spike firing rate of a sample from SC-100 dataset. Ground truth start ($t_{start}$) and end ($t_{end}$) points of keywords are indicated by yellow dashed lines, while the early ($t_{d}$) and late ($T$) decision time is denoted by a red and orange dashed line respectively.}
  \label{fig:spike rate viz}
\end{figure}

In this section, we explore the energy efficiency benefits of the early-decision SNN. As we mentioned Section \ref{sec:comparison}, our proposed ED-sKWS can reduce energy consumption. For more detailed analysis, we do visualization on a sample from SC-100 dataset with the keyword utterance ending time annotation. Figure \ref{fig:spike rate viz} show the spike firing rate in all timesteps, which is correlated with energy consumption. 

In Figure \ref{fig:spike rate viz}, at the beginning the spike firing rate remains notably low before the keyword begin time ($t_{begin}$). This observation highlights the event-driven nature of the SNN model, which saves energy by not activating spiking neurons without audio inputs. The spiking neurons start to activate at the beginning of the keyword ($t_{begin}$), marked by a yellow dashed line, leading to a noticeable increase in the spike firing rate to 0.2. At the keyword end time $t_{d}$, our ED-sKWS model has enough confidence to make the prediction, which is before the end of speech, demonstrating the rapid response. Interestingly, the spike firing rate does not immediately decrease after the keyword ends at $t_{end}$. Instead, it remains at about 0.175 for several timesteps before gradually decreasing. This is because the spiking neuron stores the historical information in the membrane potential $U$, and may still fir without the additional input. Our early-decision strategy can end the processing immediately after the early decision, significantly reducing the number of computation operations from the early-decision timestep $t_d$ to the late-decision timestep $T$, thereby enhancing the model's energy efficiency.

\section{Conclusion}
We introduce ED-sKWS, an SNN-based model designed for early decision-making in real-time Keyword Spotting (KWS) tasks. This model is capable of making predictions at each timestep.  Furthermore, we present the SC-100 dataset, featuring 100 daily life speech commands, each annotated with precise start and end timestamps to facilitate the evaluation of early decision timing. Experimental results on both Google Speech Command and SC-100 datasets reveal that our ED-sKWS model can achieve competitive performance with about 61\% timesteps and 52\% energy consumption, highlighting its capacity for rapid and energy-efficient keyword spotting.  Additionally, the proposed CT loss significantly lowers the spike firing rate, further enhancing the model's energy efficiency.

\newpage
\bibliographystyle{IEEEtran}
{\normalsize
\bibliography{mybib}

\begin{thebibliography}{10}
\providecommand{\url}[1]{#1}
\csname url@samestyle\endcsname
\providecommand{\newblock}{\relax}
\providecommand{\bibinfo}[2]{#2}
\providecommand{\BIBentrySTDinterwordspacing}{\spaceskip=0pt\relax}
\providecommand{\BIBentryALTinterwordstretchfactor}{4}
\providecommand{\BIBentryALTinterwordspacing}{\spaceskip=\fontdimen2\font plus
\BIBentryALTinterwordstretchfactor\fontdimen3\font minus \fontdimen4\font\relax}
\providecommand{\BIBforeignlanguage}[2]{{%
\expandafter\ifx\csname l@#1\endcsname\relax
\typeout{** WARNING: IEEEtran.bst: No hyphenation pattern has been}%
\typeout{** loaded for the language `#1'. Using the pattern for}%
\typeout{** the default language instead.}%
\else
\language=\csname l@#1\endcsname
\fi
#2}}
\providecommand{\BIBdecl}{\relax}
\BIBdecl

\bibitem{lopez2021deep}
I.~L{\'o}pez-Espejo, Z.-H. Tan, J.~H. Hansen, and J.~Jensen, ``Deep spoken keyword spotting: An overview,'' \emph{IEEE Access}, vol.~10, pp. 4169--4199, 2021.

\bibitem{song2022knowledge}
Z.~Song, Q.~Liu, Q.~Yang, and H.~Li, ``Knowledge distillation for in-memory keyword spotting model.'' in \emph{INTERSPEECH}, 2022, pp. 4128--4132.

\bibitem{wu2018spiking}
J.~Wu, Y.~Chua, M.~Zhang, H.~Li, and K.~C. Tan, ``A spiking neural network framework for robust sound classification,'' \emph{Frontiers in Neuroscience}, vol.~12, p. 836, 2018.

\bibitem{yin2021accurate}
B.~Yin, F.~Corradi, and S.~M. Boht{\'e}, ``Accurate and efficient time-domain classification with adaptive spiking recurrent neural networks,'' \emph{Nature Machine Intelligence}, vol.~3, no.~10, pp. 905--913, 2021.

\bibitem{liu2020unsupervised}
Q.~Liu, G.~Pan, H.~Ruan, D.~Xing, Q.~Xu, and H.~Tang, ``Unsupervised aer object recognition based on multiscale spatio-temporal features and spiking neurons,'' \emph{IEEE Transactions on Neural Networks and Learning Systems}, vol.~31, no.~12, pp. 5300--5311, 2020.

\bibitem{Liu2024120660}
Q.~Liu, M.~Ge, and H.~Li, ``Intelligent event-based lip reading word classification with spiking neural networks using spatio-temporal attention features and triplet loss,'' \emph{Information Sciences}, vol. 675, p. 120660, 2024.

\bibitem{wu2018biologically}
J.~Wu, Y.~Chua, and H.~Li, ``A biologically plausible speech recognition framework based on spiking neural networks,'' in \emph{2018 International Joint Conference on Neural Networks}.\hskip 1em plus 0.5em minus 0.4em\relax IEEE, 2018, pp. 1--8.

\bibitem{yilmaz2020deep}
E.~Y{\i}lmaz, O.~B. Gevrek, J.~Wu, Y.~Chen, X.~Meng, and H.~Li, ``Deep convolutional spiking neural networks for keyword spotting,'' in \emph{INTERSPEECH}, 2020, pp. 2557--2561.

\bibitem{yang2022deep}
Q.~Yang, Q.~Liu, and H.~Li, ``Deep residual spiking neural network for keyword spotting in low-resource settings.'' in \emph{INTERSPEECH}, 2022, pp. 3023--3027.

\bibitem{sun2023adaptive}
P.~Sun, E.~Eqlimi, Y.~Chua, P.~Devos, and D.~Botteldooren, ``Adaptive axonal delays in feedforward spiking neural networks for accurate spoken word recognition,'' in \emph{ICASSP 2023-2023 IEEE International Conference on Acoustics, Speech and Signal Processing}.\hskip 1em plus 0.5em minus 0.4em\relax IEEE, 2023, pp. 1--5.

\bibitem{song2024spiking}
Z.~Song, J.~Wu, M.~Zhang, M.~Z. Shou, and H.~Li, ``Spiking-leaf: A learnable auditory front-end for spiking neural networks,'' in \emph{ICASSP 2024-2024 IEEE International Conference on Acoustics, Speech and Signal Processing}.\hskip 1em plus 0.5em minus 0.4em\relax IEEE, 2024, pp. 226--230.

\bibitem{bellec2018long}
G.~Bellec, D.~Salaj, A.~Subramoney, R.~Legenstein, and W.~Maass, ``Long short-term memory and learning-to-learn in networks of spiking neurons,'' \emph{Advances in Neural Information Processing Systems}, vol.~31, 2018.

\bibitem{yang2024svad}
Q.~Yang, Q.~Liu, N.~Li, M.~Ge, Z.~Song, and H.~Li, ``{SVAD}: A robust, low-power, and light-weight voice activity detection with spiking neural networks,'' in \emph{ICASSP 2024-2024 IEEE International Conference on Acoustics, Speech and Signal Processing}.\hskip 1em plus 0.5em minus 0.4em\relax IEEE, 2024, pp. 221--225.

\bibitem{choi2019temporal}
S.~Choi, S.~Seo, B.~Shin, H.~Byun, M.~Kersner, B.~Kim, D.~Kim, and S.~Ha, ``Temporal convolution for real-time keyword spotting on mobile devices,'' \emph{arXiv preprint arXiv:1904.03814}, 2019.

\bibitem{speechcommandsv2}
\BIBentryALTinterwordspacing
P.~{Warden}, ``{Speech Commands: A Dataset for Limited-Vocabulary Speech Recognition},'' \emph{ArXiv e-prints}, Apr. 2018. [Online]. Available: \url{https://arxiv.org/abs/1804.03209}
\BIBentrySTDinterwordspacing

\bibitem{bittar2022surrogate}
A.~Bittar and P.~N. Garner, ``A surrogate gradient spiking baseline for speech command recognition,'' \emph{Frontiers in Neuroscience}, vol.~16, p. 865897, 2022.

\bibitem{gerstner2002spiking}
W.~Gerstner and W.~M. Kistler, \emph{Spiking neuron models: Single neurons, populations, plasticity}.\hskip 1em plus 0.5em minus 0.4em\relax Cambridge university press, 2002.

\bibitem{wu2018spatio}
Y.~Wu, L.~Deng, G.~Li, J.~Zhu, and L.~Shi, ``Spatio-temporal backpropagation for training high-performance spiking neural networks,'' \emph{Frontiers in Neuroscience}, vol.~12, p. 331, 2018.

\bibitem{deng2022temporal}
\BIBentryALTinterwordspacing
S.~Deng, Y.~Li, S.~Zhang, and S.~Gu, ``Temporal efficient training of spiking neural network via gradient re-weighting,'' in \emph{International Conference on Learning Representations}, 2022. [Online]. Available: \url{https://openreview.net/forum?id=_XNtisL32jv}
\BIBentrySTDinterwordspacing

\bibitem{meneses2022sidi}
M.~C. Meneses, R.~B. Holanda, L.~V. Peres, and G.~D. Rocha, ``Sidi kws: A large-scale multilingual dataset for keyword spotting.'' in \emph{INTERSPEECH}, 2022, pp. 4616--4620.

\bibitem{panayotov2015librispeech}
V.~Panayotov, G.~Chen, D.~Povey, and S.~Khudanpur, ``Librispeech: an asr corpus based on public domain audio books,'' in \emph{2015 IEEE International Conference on Acoustics, Speech and Signal Processing}.\hskip 1em plus 0.5em minus 0.4em\relax IEEE, 2015, pp. 5206--5210.

\bibitem{li2021differentiable}
Y.~Li, Y.~Guo, S.~Zhang, S.~Deng, Y.~Hai, and S.~Gu, ``Differentiable spike: Rethinking gradient-descent for training spiking neural networks,'' \emph{Advances in Neural Information Processing Systems}, vol.~34, pp. 23\,426--23\,439, 2021.

\bibitem{rathi2021diet}
N.~Rathi and K.~Roy, ``Diet-snn: A low-latency spiking neural network with direct input encoding and leakage and threshold optimization,'' \emph{IEEE Transactions on Neural Networks and Learning Systems}, 2021.

\bibitem{che2022differentiable}
K.~Che, L.~Leng, K.~Zhang, J.~Zhang, Q.~Meng, J.~Cheng, Q.~Guo, and J.~Liao, ``Differentiable hierarchical and surrogate gradient search for spiking neural networks,'' \emph{Advances in Neural Information Processing Systems}, vol.~35, pp. 24\,975--24\,990, 2022.

\bibitem{horowitz20141}
M.~Horowitz, ``1.1 computing's energy problem (and what we can do about it),'' in \emph{2014 IEEE International Solid-state Circuits Conference Digest of Technical Papers}.\hskip 1em plus 0.5em minus 0.4em\relax IEEE, 2014, pp. 10--14.

\bibitem{he2024msat}
X.~He, Y.~Li, D.~Zhao, Q.~Kong, and Y.~Zeng, ``Msat: biologically inspired multistage adaptive threshold for conversion of spiking neural networks,'' \emph{Neural Computing and Applications}, pp. 1--17, 2024.

\end{thebibliography}
}
\end{document}